\documentclass{mem}
\input{psfig}
\usepackage{natbib}
\usepackage{txfonts}\usepackage{balance}
\usepackage{graphicx}
\usepackage[a4paper]{hyperref}
\idline{1}{1}
\begin{document}
\def\teff{$T\rm_{eff }$}
\def\kms{$\mathrm {km s}^{-1}$}

\title{
CoRoT and the search for exoplanets.
}

   \subtitle{The Italian contribution}

\author{
E.\,Poretti\inst{1} \and
A.F.\,Lanza\inst{2} \and
C.\,Maceroni\inst{3} \and
I.\,Pagano\inst{2} \and
V.\,Ripepi\inst{4} 
          }

  \offprints{E. Poretti}

\institute{
INAF -- Osservatorio Astronomico di Brera, Via E. Bianchi 46,
I-23807 Merate, Italy.
\email{ennio.poretti@brera.inaf.it}
\and
INAF -- Osservatorio Astrofisico di Catania, Via S. Sofia 78,
I-95123 Catania, Italy. {\email nlanza,ipa@oact.inaf.it}
\and
INAF -- Osservatorio Astronomico di Roma, Via Frascati 33,
I-00040 Monte Porzio, Italy. \email{maceroni@mporzio.astro.it} 
\and
INAF -- Osservatorio Astronomico di Capodimonte, Via Moiarello 16,
I-80131 Napoli, Italy. \email{ripepi@na.astro.it}
}

\authorrunning{Poretti et al.}

\titlerunning{CoRoT and exoplanets}

\abstract{The space  mission CoRoT (COnvection, ROtation and planetary Transits) 
will offer the possibility to detect extrasolar planets by means of the transit method.
The satellite will observe about 60000 targets in the range 11.0$<V<$16.0,
located in five fields near the equator. The parts of the preparatory work  in which the
Italian community has been involved are described, with a particular emphasis
on the observations to characterize the fields, on the effects of  stellar activity
and background stars and on the outreach activities. These actions complement
those undertaken for the asteroseismologic programme.
}
\maketitle{}

\section{Introduction}

The satellite CoRoT (COnvection, ROtation and planetary Transits) will perform
photometric observations of stars with unprecedented high accuracy.
Its goal 
is twofold: the study of the stellar interiors (the asteroseismic part) and the
search for extrasolar planets (the exoplanetary part).
The launch is currently scheduled for December 2006.
The mission has a unique instrument: a 27--cm aperture telescope equipped with
two CCDs for each scientific case.
The selected CoRoT direction of pointing  is
a double--cone (the CoRoT eyes) centered at $\alpha$=6h50m/18h50m 
(galactic Anticenter/Center), $\delta=0^{\rm o}$;
the radius of each eye is 10$^{\rm o}$. 
To achieve its goals, CoRoT will uninterruptly observe five
fields for 150~d each (long runs). 
For a more detailed description of the
instrument visit the site {\tt http://smsc.cnes.fr/COROT/}.


Besides a major French
participation (80\%), the CoRoT mission also involves other european countries:
Spain, Austria, Belgium, Germany as well as a participation of ESA. 
After an initial participation which collected a wide interest in the national
community, 
the Italian Space Agency ASI withdrew the official support to the project.
Despite this, a group of researchers still remained active on
various aspects. \citet{canarie} described the Italian contribution
to the asteroseismologic part. Here we report on  the contribution to the
exoplanetary part. Four main issues are addressed: the characterization of the
stellar content of the exoplanetary fields, the disentangling of the planetary
transits from stellar activity, the influence of background stars in the 
transit detection, the preparatory tools and outreach activities.

\section{Stars and planets}
\begin{figure}
\centerline{\psfig{file=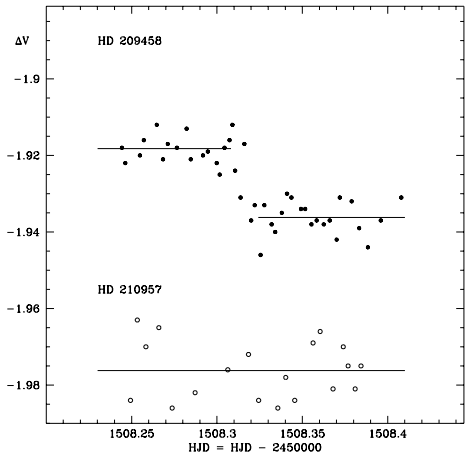,width=7.0truecm}}
\caption{\footnotesize $V$--band light curve of HD 209458 observed on the night of
 November 25, 1999  
at Merate Observatory. The transit of the planet originates with the
dimming in the luminosity of the star; the upper line finishes at the
time of the first contact, the lower one starts at the time of the
second one. The measurements
of the check star, which are arbitrarily shifted,
 are reported for the purpose of comparison. } 
\label{marcon}
\end{figure}

In the asteroseismic channel the long--duration and continuous CoRoT survey 
will result in high--precision photometry, with an expected 
noise level in the frequency spectrum of 0.7~ppm (parts per million) over a 5~d 
time baseline for a $V\sim$6.0 star. This means that  CoRoT will be able to detect 
solar--like oscillations.
In the exoplanetary channel, CoRoT will monitor simultaneously up to 12000 stars
in each long run in the 11.0$<V<$16.0 range. Thus, during the expected 2.5--y satellite lifetime,
a total of 60000 light curves will be produced at a sampling rate of 8~min. 
The detection of the extrasolar planets will be perfomed by measuring the weak decrease
of the star flux due to the transit of a planet in front of the star disk. The first
observational evidence of this phenomenon has been obtained by \citet{henry} on
HD~209458.  Figure~\ref{marcon} shows unpublished
observations of another transit of that planet, measured some weeks after its discovery 
with the 50--cm telescope at Merate  Observatory. 
For the brightest stars color information will be available and will help in discriminating single transit
events from stellar activity or other artefacts. 
To get these supplementary and useful data, a dispersive
element has been included in the optical path of the exoplanetary channel. In such a way
the point spread function of the stellar image is split roughly in three colours.

The main targets of the exoplanetary mission are bodies having the size of the Earth.
Some hypotheses have been made
to estimate the number of possible transits due to telluric planets (numbers of 
young stars with dust  discs, radii and distances of the planets from the parent
stars,~...). With a photometry optimised up to $V$=15.5 (expected precision at the
0.1 mmag level) the expected number of detections is \citep{leger}:

\begin{enumerate}
\item 25 planets having a radius of 1.6~$R_\oplus$  at 0.3~AU;
\item 40 planets having a radius of 2.0~$R_\oplus$  at 0.3~AU;
\item a few planets of around 2.0~$R_\oplus$  in the ``habitable" zone
(with the indispensable help of the chromatic information);
\item several hundreds of hot--Jupiters and Uranus--like planets, with
detailed light curves.
\end{enumerate}

The two CoRoT programmes are not independent:
once a specific asteroseismic target has been proposed (primary target), 
the corresponding exoplanetary field has to match the requirements for a fruitful
search for transits. Moreover, we also need to assure that the stars 
surrounding the primary target are also interesting for asteroseismology.
Therefore, some preliminary investigations 
of both the asteroseismologic and exoplanetary fields 
are mandatory. 
Here we report on two specific and different applications.

\subsection{The case study of HD 52265}
HD 52265 will be the target of the second CoRoT long run in the Anticenter direction
(September 2007--March 2008).  
It has been proposed for an asteroseismic investigation since
it is the only star with an already known planet in the CoRoT eyes
\citep{butler}. Transits do not occur since the planet is
not on our line of sight. However, the asteroseismic
investigation of HD~52265 should allow us a better understanding of the properties
of its planetary systems; moreover, considering
that HD~52265 has a  mass of  1.1~M$_\odot$, at the same time  
removing the observational singularity of our Solar System.
The approach
to study the parent star to investigate its planetary system
could be an important  guideline for future space missions
and perhaps HD~52265 could constitute an interesting pathfinder.

The characterization of the HD~52265 field resulted in the 
detection of several pulsators:
the $\delta$ Sct star HD~50870 \citep{poretti}, the hot Be stars
HD~50891, HD~51193, HD~51242 and the cold Be star HD~51404.
Therefore, the field has been validated
for the asteroseismic programme and looks very appealing.

\subsection{The characterization of the exoplanet fields}
In order to contribute to the characterization of the exoplanetary field
aside HD~49434 and HD~49933,  
the open cluster Dolidze 25 has been observed.
In particular, a photometric ($RI$ filters)
and spectroscopic investigation  has been carried out by using the VIRMOS@VLT
instrument \citep{virmos}. Photometry has been used to select targets for
spectroscopy. In total we obtained $\sim$900 spectra with medium
resolution (2.5~\AA/pixel) and $\sim$600 with higher resolution
(0.6~\AA/pixel). The spectra have been reduced using the ESO pipeline, and
an automatic procedure is going to be applied to obtain the
spectroscopic classification for all the investigated objects.
This, in turn, will allow us to test
the spectral types previously obtained through
photometry only. The main goal is to separate dwarf from giant stars in
the most reliable way.  After such procedure, only the dwarf stars will be
selected for additional photometric monitoring.

\begin{figure}[]
\psfig{file=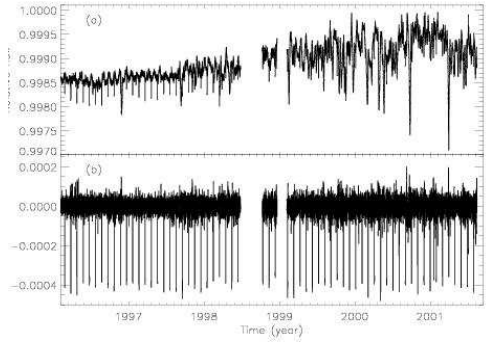,width=7.0cm,angle=0} 
\caption{\footnotesize
(a) The transit of an Earth--like planet of radius $R=2.3R_{\oplus}$ and orbital 
period of 30.0~d superposed on the total solar irradiance variation;
the irradiance values are normalized to the maximum observed in the given time interval;
(b) The same time series after subtracting the best fit
model light curve computed with the model after \citet{lanzatre}. The transit dips are
now clearly visible also during the phases with moderate or high solar activity, whereas 
they were lost in the irradiance fluctuations during the same phases in panel (a).
}
\label{figlanza1}
\end{figure}

\begin{figure*}
\unitlength1cm
\begin{minipage}[]{10cm}
\psfig{file=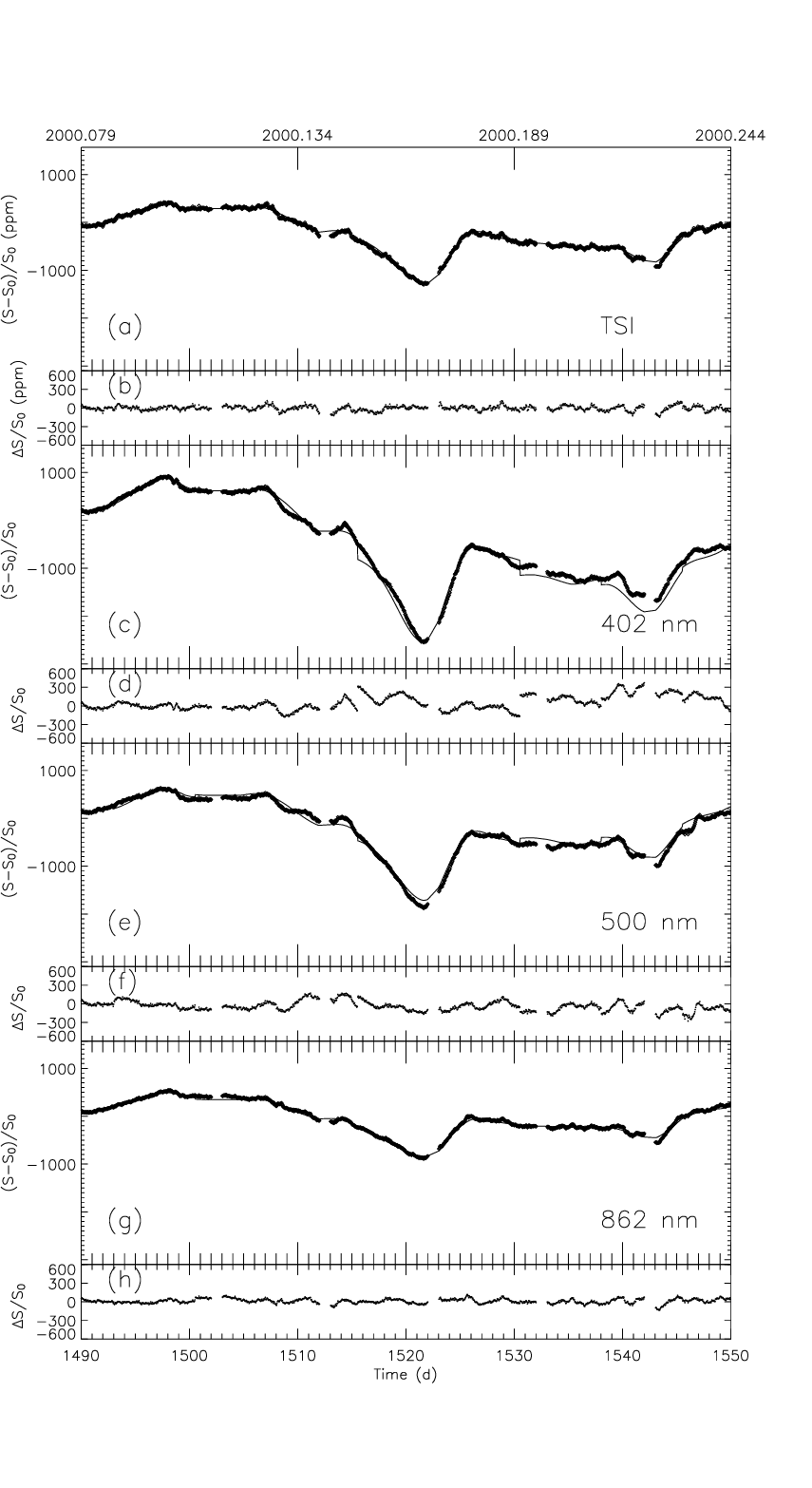,width=7.0cm}  
\end{minipage}
\hfill
\begin{minipage}[]{10cm}
\psfig{file=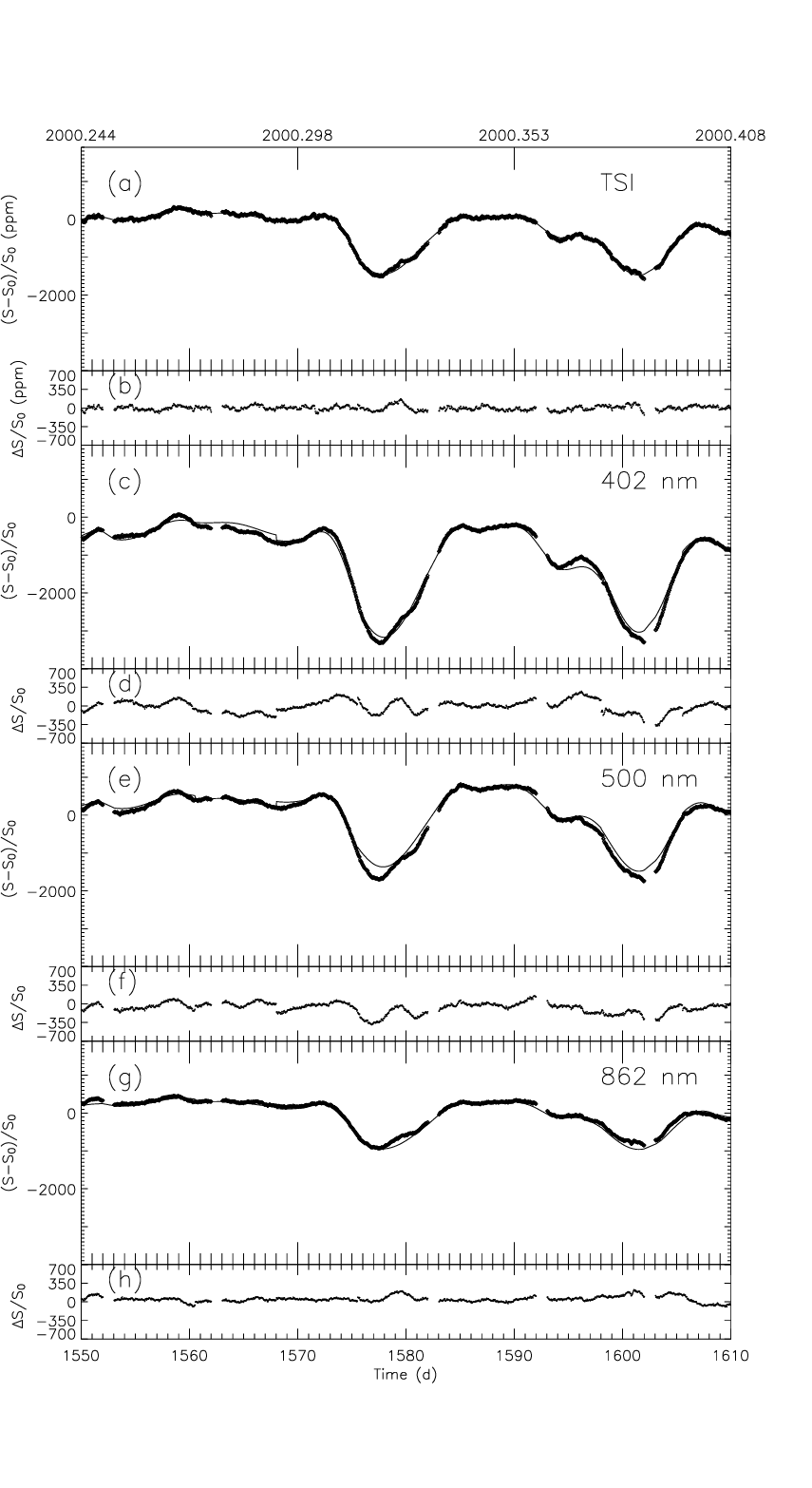,width=7.0cm}  
\end{minipage}
\caption{\footnotesize
Two subsets of the time series of the total solar irradiance (TSI) and spectral irradiances at 402, 500 and
862~nm (black dots) with their respective
best fits (solid lines) are plotted in panels (a), (c), (e) and (g) for the labelled passbands
in the left and the right columns, respectively. The best fits are obtained with
the Sun-as-a-star model introduced
by Lanza et al. (2004). The fits are usually embedded into the data sequence, except when
data gaps are present. The residuals are plotted in the corresponding lower panels
(b), (d), (f) and (h), respectively. The time is indicated in days from 1$^{\rm st}$ January 1996
on the lower scale and in years on the upper scale. The data subsets are close to the maximum of
cycle 23 and range from
29 January 2000 to 29 March 2000 in the left  panels,  and from 29 March 2000 to 28 May 2000
in the right  panels, respectively.  }
\label{figlanza2}
\end{figure*}
\begin{figure*}
\centerline{
\psfig{file=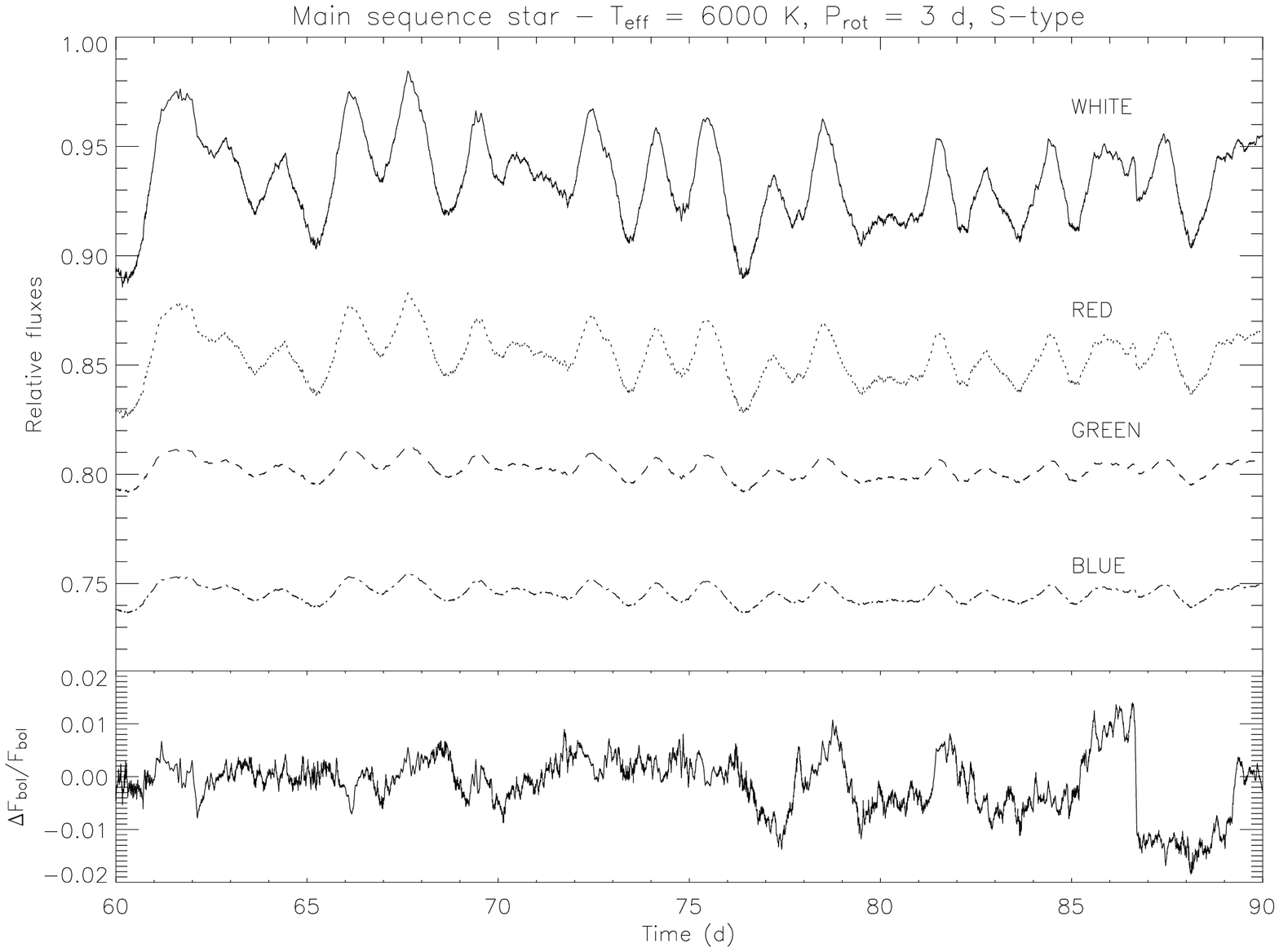,width=12.5cm,height=8.5cm,angle=90}
}
\caption{\footnotesize
{\it Top panel:} A sample of the synthetized light variations in the four CoRoT
passbands for a
main--sequence star with
$T_{\rm eff} = 6000$ K, $\log g = 4.5$ (cm s$^{-2}$), a rotation period
of 3 days and with spots dominating the light variations.
 Different linestyles refer to different passbands: solid -- white
channel; dotted -- red channel; dashed -- green channel; dash-dotted -- blue channel.
The mean values of the red, green and blue passband fluxes have been
shifted up
in order to plot them on the same scale.  {\it Lower panel}: The
short--term relative variations in the white channel that were superposed on the rotationally
modulated variations to obtain the light curve in the upper panel. The short--term
variations in the other passbands were obtained from those in the white channel
by scaling the red, green and blue channels, by factors of 0.923, 1.095, 1.199,
respectively \citep{lanzasei}.}
\label{figlanza3}
\end{figure*}

\section{Magnetic activity in late--type stars and planetary transit detection }
Main--sequence late--type stars have an outer convective envelope that,
in combination with a sufficiently fast rotation, becomes the site of hydromagnetic
dynamo action \citep{parke,rudi}.
The key parameter measuring the efficiency of the magnetic field
generation is the Rossby number $Ro$, i.e., the ratio of the rotation
period  to the convective turnover time at the base of the convection
zone. When $Ro$ is of the order of the unity, the amplification and 
modulation of magnetic fields take place and
the star displays the typical
signatures of solar--like activity. The optical flux is modulated by
cool spots and bright faculae in the photosphere which evolve on time scales
from a few hours up to tens of days. The active regions with a lifetime comparable to
or longer than the stellar rotation period produce the rotational modulation of the
stellar flux.

The observations of the solar disk-integrated irradiance show that
 the amplitude of the flux modulation
due to the surface brightness inhomogeneities is of the order of $2000-3000$ ppm 
around the maximum of the 11--yr cycle  \citep{froh}. The detection of Earth--like
planetary transits  requires
the observation of flux dips with amplitudes of the order of 100 ppm.
This implies that solar--like activity
introduces a severe limitation for the detection of Earth--like transits across the disks of stars
similar to the Sun or with a higher level of activity \citep{aigrain}.
                                                                                                                                                      
In the framework of the preparatory work for the CoRoT mission, we 
built models of the variability of the Sun as a star.
They are based on a discrete distribution of active regions plus a uniformly
distributed background whose area and coordinates vary with a time scale of 14~d
in order to reproduce the modulation of the total as well as spectral solar irradiances at
402, 500 and 862~nm \citep{lanzatre,lanzaquattro}. Such a model
allows us to fit the solar variation giving residuals of the order of 100--200 ppm for the
total irradiance which may significantly improve the detection of planetary transit 
(Figs.~\ref{figlanza1}
and \ref{figlanza2}). Moreover, the model can be applied to scale the solar variability to the case of
younger, more active stars to produce simulated light curves (Fig.~\ref{figlanza3}) which will
be used for testing different
detection algorithms for the CoRoT mission, i.e., the so-called CoRoT Exoplanet Blind Tests 1 and 2,
described by
\citet{moutou,lanzasei} and Moutou et al. (in preparation).
                                                                                                                                                      
The model described above as well as those applied to stars with a high
level of magnetic activity \citep{messinadue,lanzadue}
allow us to derive information also
on the stellar rotation period, the level of
activity and the kind of hydromagnetic dynamo operating in a given star. The level of
activity determines the average UV and X--ray stellar fluxes \citep{messinatre}
which are important to model the chemistry of  exoplanet atmospheres.
On the other hand, close--by Jupiter--sized exoplanets may affect stellar rotation and
magnetic activity through tidal as well as magnetospheric interaction \citep{saar, mcivor}.
                                                                                                                                                   
Modelling the light curves obtained by CoRoT will
 allow us to obtain information on the area variation and surface distribution
of stellar  active regions which will disclose the dependence of
magnetic activity on fundamental
stellar parameters, that is presently not known for stars with an activity level
comparable to that of the Sun. 

\section{Planet detection and background binaries}
Another very  serious problem for  planetary transit detection
will  arise from  the contamination by background
eclipsing binaries (Bebs).
The CoRoT exoplanet program essentially performs
aperture photometry on a crowded stellar field. In spite of the masks,
conceived to optimize photometry and to limit
contamination from nearby stars, a  contribution of
faint background objects to the light curve of the target is
unavoidable \citep{praga}.
The high frequency of false alarms from Bebs was already 
experienced in the OGLE planetary transit search: only five
true planets out of 177 alarms were confirmed by
follow--up observations \citep{ogle},  a result in agreement
with the theoretical estimation of \citet{brown}.
                                                                                                                                                   
We need to test the ability of detrending the lightcurves from instrumental and
environmental biases and of  eventually  detecting planets in
the specific context of CoRoT. Therefore, 
a complex simulator has been realized \citep{auve} to reproduce instrumental noises and
to produce the first light curves for the exoplanet channel. 
The simulator was used to generate a sample output of 1000 lightcurves, hiding 20 transiting planets
and 20 other types of variable objects of similar amplitude (Bebs, binaries with
grazing eclipses, other variables); such a sample was used in a first blind test (BT1) to
estimate the  detection threshold of CoRoT \citep{moutou}, i.e. the minimum detectable
planet radius  depending on  the period and parent star. It served as well to test the
efficiency of the various detrending/detection methods  developed by the different
teams who will  work on the real data.
                                                                                                                                                   
The results  of BT1 allowed to derive an empirical ``detection curve", expressing a
lower limit of transit depth for detection as $d \simeq 10^{-3} n^{-1/2}$,
where $d$ is the depth and $n$ the number of transits during the run.
For instance a planet orbiting a G0V star should be detected if $R_{pl}\geq 2-4 R_\oplus$
(for $P_{orb}=3-50$~d).
Besides, it was directly confirmed that Bebs will be in practice the only
source of false alarms: the only cases of misidentification by all teams,
and therefore not method--related, are from eclipsing binaries.

A more refined blind test (BT2)  is now in progress with the purpose of
analyzing the problem of Bebs and of estimating the type and amount of
necessary follow--up observations.  Many cases could be discriminated
just by detailed analysis of the folded lightcurve (on the basis of
color behavior of the transit and by detection of a secondary minimum).
The BT2 light curves were produced using the same complex
simulator used in BT1, but polluting events, whose
statistics has been estimated using the Corotlux tool \citep{lux},
have been included.
To provide lightcurves
in the three color ``bandpasses" of CoRoT, Corotlux uses the Besan\c {c}on
galactic model \citep{robin}, real observations of the CoRoT fields
(to estimate the number of stars of different type in each field) and
the expected binary frequency  and  distribution
of orbital elements (Moutou et al., in preparation). 
A first result is that in one long run in the CoRoT Anticenter field a
mean number of $\sim80$ candidate planets is expected, with about
80\% of alarms cleared by follow up photometry and 10\% by follow--up
spectroscopy \citep{pont}.


\begin{figure*}[]
\psfig{file=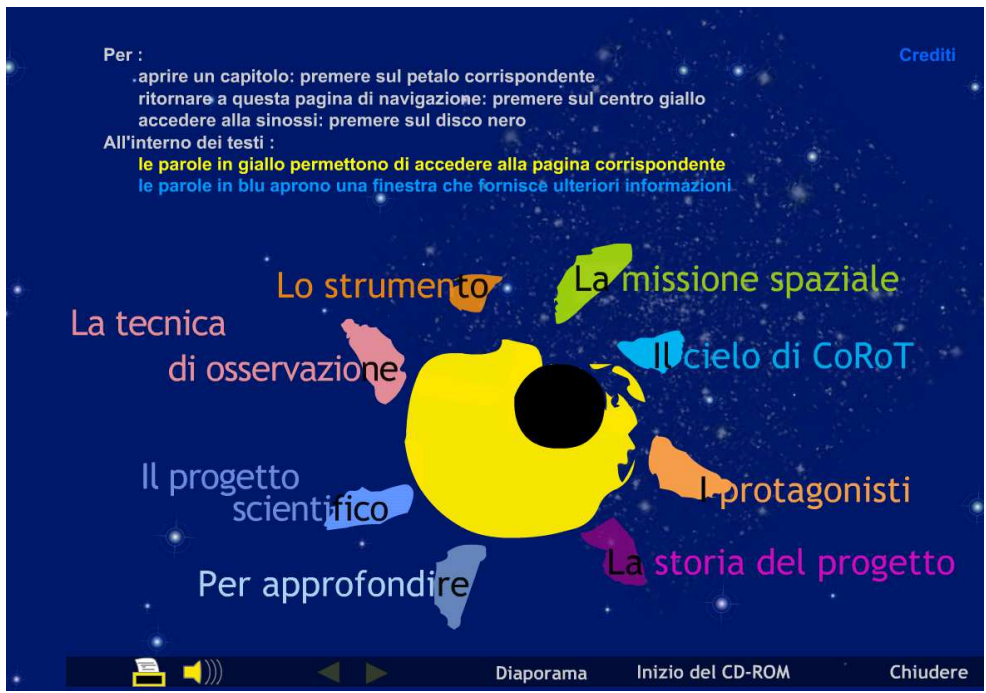,width=13cm,angle=0} 
\caption{\footnotesize
Starting menu of the "Dal cuore delle stelle  ... ai pianeti abitabili" multimedia flash project.
}
\label{pagano1}
\end{figure*}

\begin{figure*}[]
\begin{center}
\begin{minipage}[!t]{10cm}
\centerline{\psfig{file=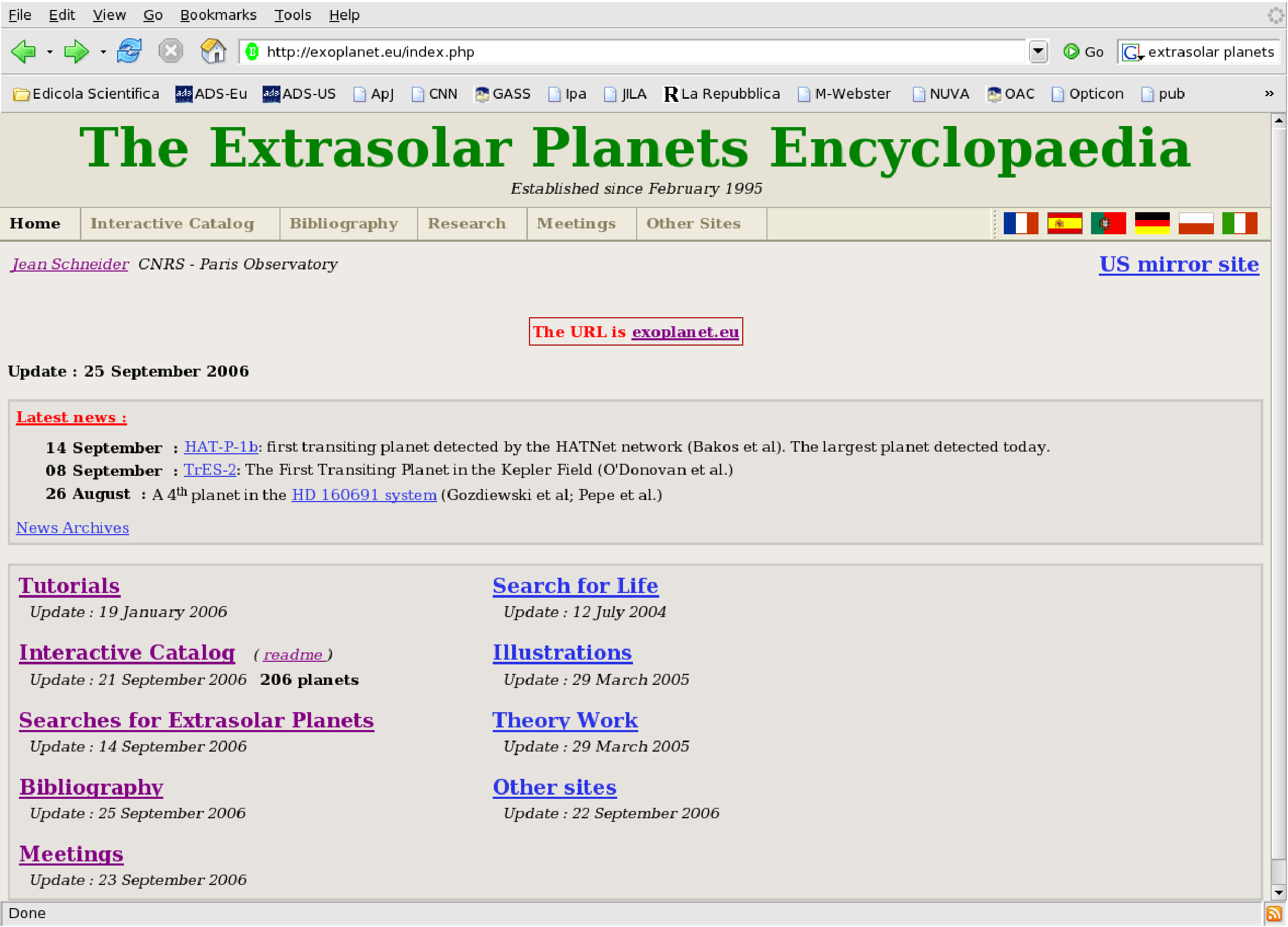,width=12cm,angle=0}} 
\end{minipage}
\hfill
\begin{minipage}[!t]{10cm}
\centerline{\psfig{file=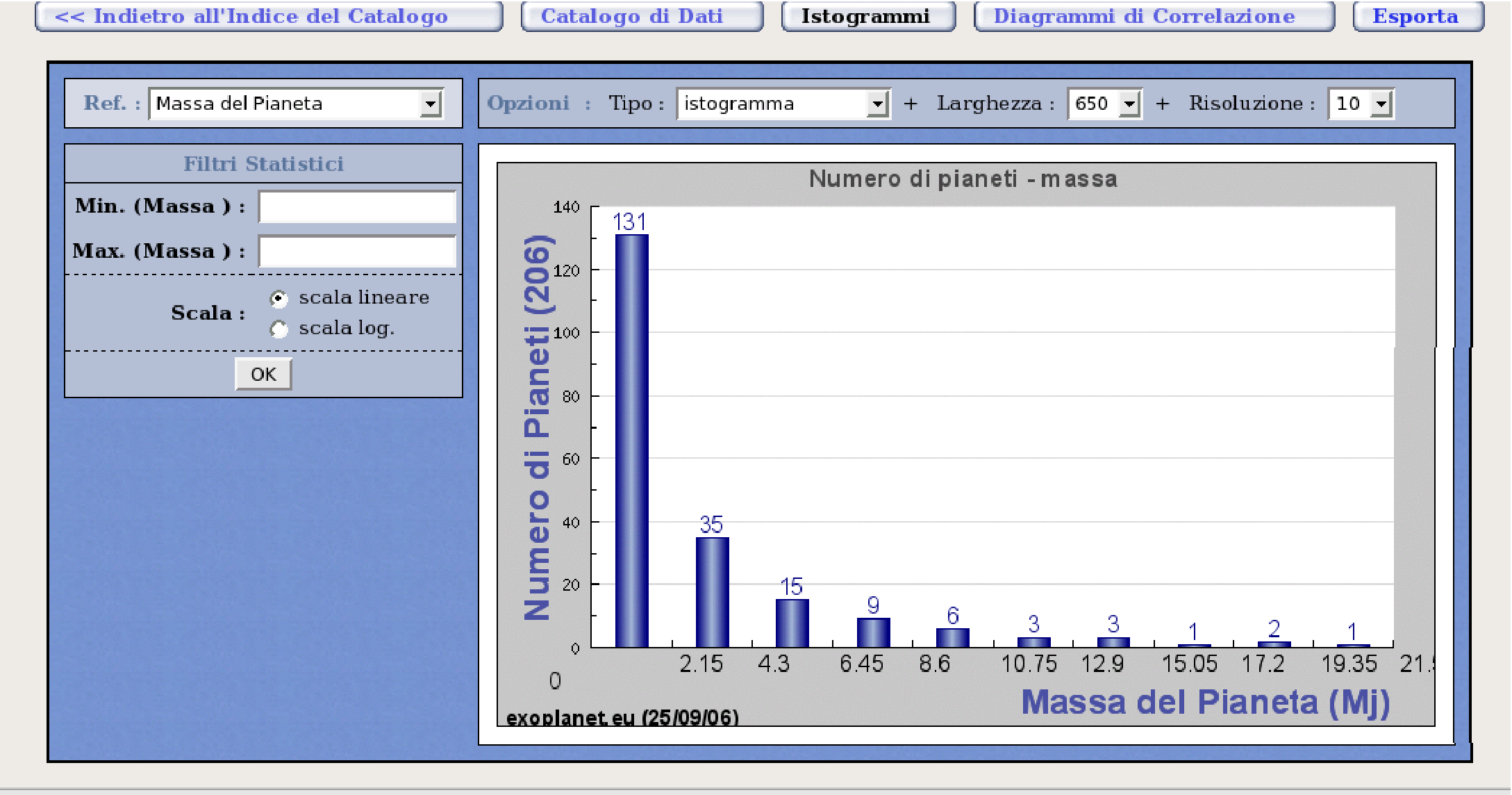,width=12cm,angle=0}}
\end{minipage}
\caption{\footnotesize {\it Top panel}: Homepage of the extrasolar planets encyclopaedia (http://exoplanet.eu).
{\it Bottom panel}: Distribution of extrasolar planet masses. This is an example of the  histograms and
correlation diagrams
that can be build from the interactive catalog of the extrasolar planets encyclopaedia.
}
\label{pagano3}
\end{center}
\end{figure*}
                                                                                                                                                   
\section{Outreach activities}
We are taking care of outreach activities related to the CoRoT mission and to the research
on extrasolar planets.
\subsection{CoRoT mission}
A multimedia product (Macromedia Flash) illustrating the CoRoT mission and its science objectives was
developed in French at the Observatoire de Paris.
We have edited and translated the project in Italian and published it on the
web site of Catania Observatory (http://www.ct.astro.it/gass/CD-CoRoT/corot.swf).
The multimedia project, titled ``Dal cuore delle stelle  ... ai pianeti abitabili''
\footnote{From the stellar core ... to habitable planets}, allows the user
to explore the following subjects (Fig.~\ref{pagano1}):
\begin{enumerate}
\item the space mission;
\item the instrument;
\item the observational technique;
\item the CoRoT sky;
\item the science project;
\item the history of the project;
\item the involved people and institutions
\item a section with background information, from basic definition 
to the physics of oscillations and stellar evolution.
\end{enumerate}
Each of the above subjects is a starting point to the exploration of several sub--subjects,
as well as it is possible to access to pop--up insights by clicking on highlighted words.
Our aim is to put this multimedia product on CD--ROM support and distribute it to schools 
and to the general public
during education and outreach activities regurarly carried out in the INAF Observatories
involved in the CoRoT projects.
\subsection{The extrasolar planets encyclopaedia}
                                                                                                                          
The extrasolar planets encyclopaedia (http://exoplanet.eu) is a working tool providing
all the latest detections and data announced by professional 
astronomers, useful to facilitate progress in exoplanetology.
It has been established and mantained by Jean Schneider of the  
CNRS -- Paris Observatory since February 1995.
From the homepage (Figure~\ref{pagano3}, top panel) it is 
possible to access an interactive catalog of all known extrasolar
planets, and to update  information on bibliography,  meetings, ongoing research programs and future
projects.  Finally, links to related sites are provided.

The catalog gives access to interactive tools to perform statistical analysis, such as the possibility to
build histograms (Fig.~\ref{pagano3}, bottom panel), and to correlate couples of different parameters.
The site can be viewed in several languages: English, French,
Spanish, Portuguese, German, Polish and Italian, allowing the general public and amateur astronomers to
easily access
 high level scientific information in the field.  
We have provided the Italian translation and collaborate to the 
continuous update of the Italian version of the site
(http://exoplanet.eu/index.it.php).

\section{Conclusions}

We showed how the Italian contribution to the CoRoT space mission 
has allowed a quantitative evaluation
of instrumental and stellar effects in the planetary transit detection
and a more careful evaluation of the stellar content of some specific
stellar fields.
Also considering 
the full spectroscopic characterisation of the targets
and the precise photometric evaluation of the stellar variability
in the CoRoT fields \citep{canarie}, 
the Italian researchers provided original and useful inputs 
to the scientific profile to the mission.
\begin{acknowledgements}
The authors wish to thank all the Italian researchers and students 
(post--doc, PhD, undergraduate)
who are participating to the CoRoT activities; they  bear witness of the large interest of our
community into this exciting and pioneering space mission. Thanks are also due to
Pierre Barge, Jean Schneider and Katrien Uytterhoeven for comments on a first draft of the manuscript.
\end{acknowledgements}

\end{document}